\documentclass{article}
\usepackage{graphicx}
\usepackage{amsmath}
\usepackage{amssymb}
\usepackage{amsfonts}
\usepackage{authblk} 

\title{Physical Selection of Cartan Soft Breakings in a Broken $SU(3)$ Yang--Mills Model}

\author[1]{M. M. Amaral\thanks{mramaciel@gmail.com}}
\author[2]{V. E. R. Lemes\thanks{verlemes@gmail.com}}

\affil[1]{\small \em Gauge Freedom, Inc., Los Angeles, CA, USA}
\affil[2]{\small \em Instituto de F\'\i sica, Universidade do Estado do Rio de Janeiro, \protect\\ Rua S\~{a}o Francisco Xavier 524, Maracan\~{a}, Rio de Janeiro - RJ, 20550-013, Brazil}

\date{\today}

\begin{document}

\maketitle

\begin{abstract}
We revisit the broken-$SU(3)$ Yang--Mills model with soft quadratic breaking induced by the dimension-two operator $d^{abc}A_\mu^bA_\mu^c$, previously studied for a physical source aligned along the color-3 direction. We extend the construction to a generic Cartan-valued source and determine the corresponding quadratic propagators and their associated $i$-particle structure. We show that the diagonal $(3,8)$ sector always reorganizes into a conjugate pair of complex modes and therefore supports a physical one-loop composite channel with a K\"all\'en--Lehmann spectral representation. By contrast, the replica-like off-diagonal channel built from the $(4,5)$ and $(6,7)$ sectors survives as an admissible physical channel only on a special locus in the Cartan plane. In this sense, the original color-3 model is not merely a convenient example: it is dynamically singled out by the simultaneous survival of the neutral off-diagonal physical channel together with the diagonal one. The resulting picture provides a sharper understanding of the relation between soft color breaking, confinement-inspired propagators, and the construction of physically meaningful composite operators in confining gauge theories. Furthermore, we demonstrate that this distinguished soft-breaking structure can be dynamically generated via a BRST-invariant spontaneous symmetry breaking mechanism that preserves the strict algebraic renormalizability of the theory.
\end{abstract}

\section{Introduction}

The problem of gluon confinement remains one of the central questions in the study of strongly coupled gauge theories. From the point of view of the elementary propagators, an important lesson emerging from both analytical and numerical investigations is that the infrared behavior of the gluon field differs drastically from that of an ordinary particle excitation. In particular, Gribov-like propagators display complex poles and violate positivity, thereby obstructing a particle interpretation for the elementary gauge field \cite{Gribov:1977wm,Zwanziger:1991ac,Zwanziger:1992qr,Dudal:2008sp}. At the same time, suitable local composite operators may still possess a K\"all\'en--Lehmann representation with positive spectral density, so that physically meaningful channels can survive at the level of neutral observables \cite{Baulieu:2009ha,Sorella:2010it}.

In previous work \cite{Amaral:2013ReplicaSU3}, we studied a broken-$SU(3)$ Yang--Mills model in the Landau gauge in which a soft dimension-two breaking proportional to $d^{abc}A_\mu^bA_\mu^c$ induces a replica-like structure without the introduction of auxiliary fields. For a physical source aligned along the color-3 direction, the quadratic theory organizes itself in a remarkable way: two mixed $SU(2)$ subsectors become related by a mirror-like $i$-particle structure, while the diagonal $(3,8)$ sector furnishes a second conjugate pair of complex modes. In that setting one can construct neutral Hermitian bilinears whose one-loop two-point functions exhibit K\"all\'en--Lehmann spectral representations with positive spectral density. The original construction therefore provided an explicit mechanism by which soft color breaking and confinement-inspired propagators could coexist with physically meaningful composite channels. More recently, related questions about the definition of physical observables in broken or complexified gauge theories have been investigated in a BRST framework, showing that the admissible observable sector depends sensitively on the symmetry-breaking realization \cite{Amaral:2020Complex,Amaral:2023Observables}.

A natural question is whether the color-3 choice is merely a convenient representative inside a broader family of equivalent Cartan breakings, or whether it is in fact a distinguished locus selected by the existence of the physical composite operators themselves. This is the question we address in the present work.

To this end, we extend the physical source from the special color-3 choice to a generic Cartan-valued source in the $SU(3)$ algebra. This is the minimal deformation of the original model, but it already leads to a nontrivial restructuring of the infrared theory. As we shall see, the different color sectors respond in qualitatively different ways to this deformation.

Our main result is that the diagonal $(3,8)$ sector is robust under the full Cartan deformation: its quadratic form always diagonalizes into a conjugate pair of complex modes, and the corresponding neutral bilinear continues to define a natural physical one-loop channel. By contrast, the off-diagonal replica-like channel built from the $(4,5)$ and $(6,7)$ sectors is much more rigid. For a generic Cartan source, these sectors no longer organize themselves into a conjugate pair with the structure required for the standard $i$-particle construction. As a consequence, the neutral Hermitian bilinear that was admissible in the original model survives only on a special locus in the Cartan plane.

This result gives a new interpretation to the old construction. The original color-3 model is not simply one representative among many equivalent soft breakings. Rather, it occupies a distinguished position in Cartan space: it is precisely the locus on which the replica-like off-diagonal channel and the diagonal physical channel coexist. In this sense, the admissibility of the composite operators furnishes a physical selection principle for the orientation of the soft source in the Cartan subalgebra.

The paper is organized as follows. In Sect.~\ref{sec:color3} we revisit the original color-3 model and isolate the ingredients that will be generalized. In Sect.~\ref{sec:cartan} we determine the quadratic propagators for a generic Cartan-valued source. In Sect.~\ref{sec:operators} we analyze the neutral composite operators and establish the survival criterion for the physical channels. In Sect.~\ref{sec:phase} we describe the corresponding phase structure in the Cartan plane. Sect.~\ref{sec:conclusions} contains our conclusions. Finally, in Appendix \ref{app:ssb}, we present a BRST-invariant spontaneous symmetry breaking mechanism that dynamically generates the Cartan soft source and establishes the complete algebraic renormalizability of the model.

\section{The original color-3 model revisited}
\label{sec:color3}

We begin by recalling the special case studied in \cite{Amaral:2013ReplicaSU3}, namely the physical source aligned along the color-3 direction,
\begin{equation}
\langle J^a\rangle=m^2\delta^{a3},
\qquad m^2>0.
\label{eq:J-color3}
\end{equation}
At quadratic order, the gauge-field sector is
\begin{equation}
S_{\mathrm{quad}}
=
\frac12\int d^4x\,A_\mu^a(-\partial^2)A_\mu^a
+
\frac{i}{2}\int d^4x\,\langle J^a\rangle d^{abc}A_\mu^bA_\mu^c.
\label{eq:Squad-color3}
\end{equation}
Using the relevant symmetric coefficients
\begin{equation}
d^{338}=\frac{1}{\sqrt{3}},
\qquad
d^{344}=d^{355}=\frac12,
\qquad
d^{366}=d^{377}=-\frac12,
\label{eq:relevant-d-color3}
\end{equation}
the quadratic action decomposes into three distinct sectors. We denote by
\begin{equation}
\theta_{\mu\nu}(k)=\delta_{\mu\nu}-\frac{k_\mu k_\nu}{k^2}
\label{eq:transverse-projector}
\end{equation}
the transverse projector in momentum space.

The $(1,2)$ sector remains massless,
\begin{equation}
\langle A_\mu^1(k)A_\nu^1(-k)\rangle
=
\langle A_\mu^2(k)A_\nu^2(-k)\rangle
=
\frac{1}{k^2}\,\theta_{\mu\nu}(k),
\label{eq:prop12-color3}
\end{equation}
while the $(4,5)$ and $(6,7)$ sectors acquire opposite imaginary mass shifts,
\begin{equation}
\langle A_\mu^4(k)A_\nu^4(-k)\rangle
=
\langle A_\mu^5(k)A_\nu^5(-k)\rangle
=
\frac{1}{k^2+i\,m^2/2}\,\theta_{\mu\nu}(k),
\label{eq:prop45-color3}
\end{equation}
\begin{equation}
\langle A_\mu^6(k)A_\nu^6(-k)\rangle
=
\langle A_\mu^7(k)A_\nu^7(-k)\rangle
=
\frac{1}{k^2-i\,m^2/2}\,\theta_{\mu\nu}(k).
\label{eq:prop67-color3}
\end{equation}
Thus the two mixed off-diagonal subsectors form a conjugate pair of the standard $i$-particle type.

The diagonal $(3,8)$ sector is mixed already at quadratic level. Its action is
\begin{equation}
S_{(3,8)}
=
\int d^4x
\left\{
\frac12 A_\mu^3(-\partial^2)A_\mu^3
+
\frac12 A_\mu^8(-\partial^2)A_\mu^8
+
i\frac{m^2}{\sqrt{3}}A_\mu^3A_\mu^8
\right\},
\label{eq:S38-color3}
\end{equation}
which yields
\begin{equation}
\langle A_\mu^3(k)A_\nu^3(-k)\rangle
=
\langle A_\mu^8(k)A_\nu^8(-k)\rangle
=
\frac{k^2}{k^4+m^4/3}\,\theta_{\mu\nu}(k),
\label{eq:prop33-88-color3}
\end{equation}
\begin{equation}
\langle A_\mu^3(k)A_\nu^8(-k)\rangle
=
\langle A_\mu^8(k)A_\nu^3(-k)\rangle
=
-\,i\,\frac{m^2/\sqrt{3}}{k^4+m^4/3}\,\theta_{\mu\nu}(k).
\label{eq:prop38-color3}
\end{equation}
Introducing the rotated fields
\begin{equation}
U_\mu=\frac{1}{\sqrt{2}}(A_\mu^3+A_\mu^8),
\qquad
V_\mu=\frac{1}{\sqrt{2}}(-A_\mu^3+A_\mu^8),
\label{eq:UV-color3}
\end{equation}
the diagonal block takes the form
\begin{equation}
S_{(3,8)}
=
\frac12\int d^4x
\left[
U_\mu\left(-\partial^2+i\frac{m^2}{\sqrt{3}}\right)U_\mu
+
V_\mu\left(-\partial^2-i\frac{m^2}{\sqrt{3}}\right)V_\mu
\right].
\label{eq:S38diag-color3}
\end{equation}
Hence the $(3,8)$ sector also reorganizes into a conjugate $i$-particle pair, now with gap $m^2/\sqrt{3}$.

The important point for what follows is that the color-3 model supports two distinct neutral one-loop channels with acceptable analytic behavior. For the off-diagonal sectors, we define
\begin{equation}
E_{\mu\nu}^{3\pm}
=
\frac{1}{\sqrt{2}}\left(F_{\mu\nu}^4\pm iF_{\mu\nu}^5\right),
\qquad
E_{\mu\nu}^{2\pm}
=
\frac{1}{\sqrt{2}}\left(F_{\mu\nu}^6\pm iF_{\mu\nu}^7\right).
\label{eq:Epm-color3}
\end{equation}
Under the residual generator $h_1=\lambda^3$, these obey
\begin{equation}
[h_1,E^{2\pm}]=\mp\frac12 E^{2\pm},
\qquad
[h_1,E^{3\pm}]=\pm\frac12 E^{3\pm},
\label{eq:charges-color3}
\end{equation}
so that the bilinear
\begin{equation}
\phi
=
E_{\mu\nu}^{2+}E_{\mu\nu}^{3+}
+
E_{\mu\nu}^{3-}E_{\mu\nu}^{2-}
=
F_{\mu\nu}^4F_{\mu\nu}^6-F_{\mu\nu}^5F_{\mu\nu}^7
\label{eq:phi-color3}
\end{equation}
is neutral and Hermitian. Because it is built from the conjugate pair carried by the $(4,5)$ and $(6,7)$ propagators, its one-loop two-point function admits the standard K\"all\'en--Lehmann construction \cite{Baulieu:2009ha,Sorella:2010it}.

The second channel comes from the diagonalized $(3,8)$ sector and is described by
\begin{equation}
\chi
=
(\partial_\mu U_\nu-\partial_\nu U_\mu)
(\partial_\mu V_\nu-\partial_\nu V_\mu).
\label{eq:chi-color3}
\end{equation}
Since $U_\mu$ and $V_\mu$ also form a conjugate pair, the corresponding one-loop correlator again has the standard $i$-particle structure.

Accordingly, the one-loop thresholds of the two physical channels are
\begin{equation}
\tau_\phi=m^2,
\qquad
\tau_\chi=\frac{2m^2}{\sqrt{3}}.
\label{eq:thresholds-color3}
\end{equation}
The original color-3 model is therefore characterized by the simultaneous coexistence of a replica-like off-diagonal physical channel and a diagonal physical channel. This is the benchmark configuration that we now deform by allowing the source to move in the full Cartan plane.

\section{Quadratic propagators for a generic Cartan-valued source}
\label{sec:cartan}

We now deform the special color-3 source of the previous section to a generic Cartan-valued source,
\begin{equation}
\langle J^a\rangle=\alpha\,\delta^{a3}+\beta\,\delta^{a8},
\qquad
\alpha,\beta\in\mathbb{R}.
\label{eq:J-cartan}
\end{equation}
At quadratic order, the gauge-field sector is
\begin{equation}
S_{\mathrm{quad}}
=
\frac12\int d^4x\,A_\mu^a(-\partial^2)A_\mu^a
+
\frac{i}{2}\int d^4x\,\langle J^a\rangle d^{abc}A_\mu^bA_\mu^c.
\label{eq:Squad-cartan}
\end{equation}
To determine the propagators it is enough to use the nonvanishing symmetric coefficients
\begin{equation}
d^{118}=d^{228}=d^{338}=\frac{1}{\sqrt{3}},
\qquad
d^{888}=-\frac{1}{\sqrt{3}},
\label{eq:d118}
\end{equation}
\begin{equation}
d^{448}=d^{558}=d^{668}=d^{778}=-\frac{1}{2\sqrt{3}},
\label{eq:d448}
\end{equation}
\begin{equation}
d^{344}=d^{355}=-d^{366}=-d^{377}=\frac12,
\label{eq:d344}
\end{equation}
together with the complete symmetry of $d^{abc}$.

It is convenient to discuss the quadratic form sector by sector. For the $(1,2)$ subsector one finds
\begin{equation}
S_{(1,2)}
=
\frac12\int d^4x
\left[
A_\mu^1\left(-\partial^2+i\frac{\beta}{\sqrt{3}}\right)A_\mu^1
+
A_\mu^2\left(-\partial^2+i\frac{\beta}{\sqrt{3}}\right)A_\mu^2
\right].
\label{eq:S12-cartan}
\end{equation}
Therefore the corresponding transverse propagators are
\begin{equation}
\langle A_\mu^1(k)A_\nu^1(-k)\rangle
=
\langle A_\mu^2(k)A_\nu^2(-k)\rangle
=
\frac{1}{k^2+i\beta/\sqrt{3}}\,\theta_{\mu\nu}(k).
\label{eq:prop12-cartan}
\end{equation}
In particular, the $(1,2)$ sector is massless only on the special line $\beta=0$.

For the $(4,5)$ subsector one obtains
\begin{equation}
S_{(4,5)}
=
\frac12\int d^4x
\left[
A_\mu^4\left(-\partial^2+i\mu_{+}^{2}\right)A_\mu^4
+
A_\mu^5\left(-\partial^2+i\mu_{+}^{2}\right)A_\mu^5
\right],
\label{eq:S45-cartan}
\end{equation}
where
\begin{equation}
\mu_{+}^{2}
=
\frac{\alpha}{2}-\frac{\beta}{2\sqrt{3}}.
\label{eq:muplus}
\end{equation}
Hence
\begin{equation}
\langle A_\mu^4(k)A_\nu^4(-k)\rangle
=
\langle A_\mu^5(k)A_\nu^5(-k)\rangle
=
\frac{1}{k^2+i\mu_{+}^{2}}\,\theta_{\mu\nu}(k).
\label{eq:prop45-cartan}
\end{equation}

Similarly, for the $(6,7)$ subsector one finds
\begin{equation}
S_{(6,7)}
=
\frac12\int d^4x
\left[
A_\mu^6\left(-\partial^2-i\mu_{-}^{2}\right)A_\mu^6
+
A_\mu^7\left(-\partial^2-i\mu_{-}^{2}\right)A_\mu^7
\right],
\label{eq:S67-cartan}
\end{equation}
with
\begin{equation}
\mu_{-}^{2}
=
\frac{\alpha}{2}+\frac{\beta}{2\sqrt{3}}.
\label{eq:muminus}
\end{equation}
Therefore
\begin{equation}
\langle A_\mu^6(k)A_\nu^6(-k)\rangle
=
\langle A_\mu^7(k)A_\nu^7(-k)\rangle
=
\frac{1}{k^2-i\mu_{-}^{2}}\,\theta_{\mu\nu}(k).
\label{eq:prop67-cartan}
\end{equation}

At this stage an important difference with the original color-3 model already becomes manifest. In the special case $\beta=0$, one has $\mu_{+}^{2}=\mu_{-}^{2}=\alpha/2$, so that the $(4,5)$ and $(6,7)$ sectors form a conjugate pair of the standard $i$-particle type. For generic $\beta\neq0$, however,
\begin{equation}
\mu_{+}^{2}\neq\mu_{-}^{2},
\label{eq:muplusminus-neq}
\end{equation}
and the replica-like mirror structure of the two off-diagonal subsectors is lost.

The remaining nontrivial block is the $(3,8)$ sector. In this case the quadratic action reads
\begin{equation}
S_{(3,8)}
=
\int d^4x
\left\{
\frac12 A_\mu^3\left(-\partial^2+i\frac{\beta}{\sqrt{3}}\right)A_\mu^3
+
\frac12 A_\mu^8\left(-\partial^2-i\frac{\beta}{\sqrt{3}}\right)A_\mu^8
+
i\frac{\alpha}{\sqrt{3}}A_\mu^3A_\mu^8
\right\}.
\label{eq:S38-cartan}
\end{equation}
In momentum space, the corresponding quadratic form is
\begin{equation}
\mathcal{M}_{(3,8)}(k)
=
\begin{pmatrix}
k^2+i\beta/\sqrt{3} & i\alpha/\sqrt{3} \\
i\alpha/\sqrt{3} & k^2-i\beta/\sqrt{3}
\end{pmatrix}.
\label{eq:M38-cartan}
\end{equation}
Its determinant is
\begin{equation}
\det\mathcal{M}_{(3,8)}(k)
=
k^4+\frac{\alpha^2+\beta^2}{3}.
\label{eq:det38-cartan}
\end{equation}
Hence the propagators in the original basis are
\begin{equation}
\langle A_\mu^3(k)A_\nu^3(-k)\rangle
=
\frac{k^2-i\beta/\sqrt{3}}{k^4+\frac{\alpha^2+\beta^2}{3}}\,
\theta_{\mu\nu}(k),
\label{eq:prop33-cartan}
\end{equation}
\begin{equation}
\langle A_\mu^8(k)A_\nu^8(-k)\rangle
=
\frac{k^2+i\beta/\sqrt{3}}{k^4+\frac{\alpha^2+\beta^2}{3}}\,
\theta_{\mu\nu}(k),
\label{eq:prop88-cartan}
\end{equation}
\begin{equation}
\langle A_\mu^3(k)A_\nu^8(-k)\rangle
=
\langle A_\mu^8(k)A_\nu^3(-k)\rangle
=
-\,i\,\frac{\alpha/\sqrt{3}}{k^4+\frac{\alpha^2+\beta^2}{3}}\,
\theta_{\mu\nu}(k).
\label{eq:prop38-cartan}
\end{equation}

The matrix (\ref{eq:M38-cartan}) is diagonalized by a real orthogonal rotation. Writing
\begin{equation}
\begin{pmatrix}
U_\mu \\ V_\mu
\end{pmatrix}
=
R(\vartheta)
\begin{pmatrix}
A_\mu^3 \\ A_\mu^8
\end{pmatrix},
\qquad
R(\vartheta)
=
\begin{pmatrix}
\cos\vartheta & \sin\vartheta \\
-\sin\vartheta & \cos\vartheta
\end{pmatrix},
\label{eq:UV-rotation-cartan}
\end{equation}
with angle determined by
\begin{equation}
\cos(2\vartheta)=\frac{\beta}{\sqrt{\alpha^2+\beta^2}},
\qquad
\sin(2\vartheta)=\frac{\alpha}{\sqrt{\alpha^2+\beta^2}},
\label{eq:theta-cartan}
\end{equation}
the diagonal sector becomes
\begin{equation}
S_{(3,8)}
=
\frac12\int d^4x
\left[
U_\mu\left(-\partial^2+i\mu_{\chi}^{2}\right)U_\mu
+
V_\mu\left(-\partial^2-i\mu_{\chi}^{2}\right)V_\mu
\right],
\label{eq:S38diag-cartan}
\end{equation}
where
\begin{equation}
\mu_{\chi}^{2}
=
\sqrt{\frac{\alpha^2+\beta^2}{3}}.
\label{eq:muchi-cartan}
\end{equation}
Thus the $(3,8)$ block always reorganizes into a conjugate pair of $i$-particles, independently of the direction of the source in the Cartan plane.

Collecting the results, the generic Cartan deformation has a very simple interpretation at quadratic level. The diagonal $(3,8)$ sector preserves its conjugate $i$-particle structure throughout the whole Cartan plane, whereas the off-diagonal sectors $(4,5)$ and $(6,7)$ form a conjugate pair only on the special line $\beta=0$. Therefore the source
\begin{equation}
\langle J^a\rangle=m^2\delta^{a3}
\label{eq:distinguished-color3}
\end{equation}
is not merely a convenient representative of the Cartan family. It is the distinguished locus on which the replica-like pairing of the off-diagonal subsectors survives.

In the next section we examine how this fact controls the existence of neutral Hermitian composite operators whose one-loop correlation functions retain the physical K\"all\'en--Lehmann structure.

\section{Neutral composite operators and the survival of the physical channel}
\label{sec:operators}

We now turn to the composite sector. The main question is whether the neutral physical channel identified in the original color-3 model survives when the source is deformed to a generic point of the Cartan plane.

As in the undeformed case, we introduce the complex field-strength combinations
\begin{equation}
X_{\mu\nu}^{\pm}
=
\frac{1}{\sqrt{2}}\left(F_{\mu\nu}^4 \pm i F_{\mu\nu}^5\right),
\qquad
Y_{\mu\nu}^{\pm}
=
\frac{1}{\sqrt{2}}\left(F_{\mu\nu}^6 \pm i F_{\mu\nu}^7\right).
\label{eq:XYpm}
\end{equation}
Under the residual generator $h_1=\lambda^3$, these satisfy
\begin{equation}
[h_1,X^{\pm}] = \pm \frac12 X^{\pm},
\qquad
[h_1,Y^{\pm}] = \mp \frac12 Y^{\pm}.
\label{eq:XYcharges}
\end{equation}
Accordingly, the four natural local Hermitian bilinears are
\begin{equation}
\Phi_1 = X^+Y^+ + X^-Y^-,
\qquad
\Phi_2 = X^+Y^- + X^-Y^+,
\label{eq:Phi12}
\end{equation}
\begin{equation}
\Phi_3 = i(X^+Y^+ - X^-Y^-),
\qquad
\Phi_4 = i(X^+Y^- - X^-Y^+).
\label{eq:Phi34}
\end{equation}
Among these, the bilinears $\Phi_1$ and $\Phi_3$ are neutral under $h_1$, while $\Phi_2$ and $\Phi_4$ are not neutral under $h_1$. In particular,
\begin{equation}
\Phi_1
=
F_{\mu\nu}^4F_{\mu\nu}^6 - F_{\mu\nu}^5F_{\mu\nu}^7,
\label{eq:Phi1real}
\end{equation}
which is precisely the off-diagonal neutral observable singled out in the original color-3 model \cite{Amaral:2013ReplicaSU3}.

However, neutrality is only a necessary condition. For a one-loop two-point function to admit the standard K\"all\'en--Lehmann interpretation of the $i$-particle type, the two constituent sectors must form a genuine conjugate pair. In the present case, the quadratic propagators found in Sect.~\ref{sec:cartan} are
\begin{equation}
\langle A_\mu^{4}(k)A_\nu^{4}(-k)\rangle
=
\langle A_\mu^{5}(k)A_\nu^{5}(-k)\rangle
=
\frac{1}{k^2+i\mu_+^2}\,\theta_{\mu\nu}(k),
\label{eq:prop45-again}
\end{equation}
\begin{equation}
\langle A_\mu^{6}(k)A_\nu^{6}(-k)\rangle
=
\langle A_\mu^{7}(k)A_\nu^{7}(-k)\rangle
=
\frac{1}{k^2-i\mu_-^2}\,\theta_{\mu\nu}(k),
\label{eq:prop67-again}
\end{equation}
with
\begin{equation}
\mu_+^2=\frac{\alpha}{2}-\frac{\beta}{2\sqrt{3}},
\qquad
\mu_-^2=\frac{\alpha}{2}+\frac{\beta}{2\sqrt{3}}.
\label{eq:muplusminus-again}
\end{equation}

The crucial point is that the analytic structure of the one-loop bubble is controlled by the location of its branch point. In the standard $i$-particle construction we restrict the off-diagonal discussion to the region in which the two parameters $\mu_+^2$ and $\mu_-^2$ are both positive, namely
\begin{equation}
\mu_+^2>0,
\qquad
\mu_-^2>0,
\label{eq:physical-region}
\end{equation}
or equivalently
\begin{equation}
\alpha>\frac{|\beta|}{\sqrt{3}}.
\label{eq:physical-region-ab}
\end{equation}
In this region, for a pair of complex mass-squared parameters $i\mu_+^2$ and $-i\mu_-^2$, the corresponding threshold is
\begin{equation}
\tau_{\mathrm{off}}
=
\left(\sqrt{i\mu_+^2}+\sqrt{-\,i\mu_-^2}\right)^2
=
2\sqrt{\mu_+^2\mu_-^2}
+
i\left(\mu_+^2-\mu_-^2\right),
\label{eq:tauoff}
\end{equation}
where principal roots are understood. A standard physical K\"all\'en--Lehmann representation requires the threshold to lie on the positive real axis. Therefore,
\begin{equation}
\Im \tau_{\mathrm{off}} = 0
\qquad\Longleftrightarrow\qquad
\mu_+^2=\mu_-^2
\qquad\Longleftrightarrow\qquad
\beta=0.
\label{eq:beta0criterion}
\end{equation}
Hence the off-diagonal replica-like physical channel survives if and only if the source lies on the color-3 line.

When $\beta=0$, one has
\begin{equation}
\mu_+^2=\mu_-^2=\frac{\alpha}{2},
\label{eq:equalmu}
\end{equation}
so that the branch point becomes
\begin{equation}
\tau_{\mathrm{off}}=\alpha.
\label{eq:tauoffspecial}
\end{equation}
This reproduces the threshold of the original model upon setting $\alpha=m^2$. In this sense, the old observable $\phi=\Phi_1$ survives exactly on the same locus on which the two off-diagonal subsectors remain a conjugate pair.

Let us now compare this behavior with the diagonal channel. From Sect.~\ref{sec:cartan}, the $(3,8)$ block always diagonalizes into two normal modes $U_\mu$ and $V_\mu$ with propagators
\begin{equation}
\langle U_\mu(k)U_\nu(-k)\rangle
=
\frac{1}{k^2+i\mu_\chi^2}\,\theta_{\mu\nu}(k),
\qquad
\langle V_\mu(k)V_\nu(-k)\rangle
=
\frac{1}{k^2-i\mu_\chi^2}\,\theta_{\mu\nu}(k),
\label{eq:UVprops-again}
\end{equation}
where
\begin{equation}
\mu_\chi^2=\sqrt{\frac{\alpha^2+\beta^2}{3}}.
\label{eq:muchi-again}
\end{equation}
Therefore the bilinear
\begin{equation}
\chi
=
(\partial_\mu U_\nu-\partial_\nu U_\mu)
(\partial_\mu V_\nu-\partial_\nu V_\mu)
\label{eq:chi-again}
\end{equation}
is always built from a genuine conjugate $i$-particle pair. Its one-loop threshold is
\begin{equation}
\tau_\chi
=
\left(\sqrt{i\mu_\chi^2}+\sqrt{-\,i\mu_\chi^2}\right)^2
=
2\mu_\chi^2
=
2\sqrt{\frac{\alpha^2+\beta^2}{3}},
\label{eq:tauchi}
\end{equation}
which is real and positive throughout the whole Cartan plane.

We may summarize the situation as follows. The diagonal physical channel $\chi$ is robust under arbitrary Cartan deformations of the source. By contrast, the replica-like off-diagonal channel survives only on the special line $\beta=0$. Thus the coexistence of the two physical channels is not generic: it is a special property of the original color-3 model.

This observation provides the first genuinely physical selection criterion for the orientation of the soft source. The point is not merely that the quadratic spectrum changes as one moves in the Cartan plane, but that the space of admissible neutral composite operators changes as well. In particular, the original source aligned along color 3 is precisely the locus for which the diagonal channel and the off-diagonal replica-like channel coexist within the standard $i$-particle construction.

\section{Phase structure in the Cartan plane}
\label{sec:phase}

The Cartan source space is not homogeneous once one asks for survival of physical channels.
The results of the previous sections allow us to describe the model in terms of a simple phase structure in the Cartan plane parametrized by the source components $(\alpha,\beta)$. Although the quadratic action is well defined for any real $\alpha$ and $\beta$, the physical content of the composite sector depends sharply on the direction of the source.

At quadratic order, the off-diagonal subsectors are governed by the effective parameters
\begin{equation}
\mu_+^2=\frac{\alpha}{2}-\frac{\beta}{2\sqrt{3}},
\qquad
\mu_-^2=\frac{\alpha}{2}+\frac{\beta}{2\sqrt{3}},
\label{eq:phase-mupm}
\end{equation}
while the diagonal $(3,8)$ block is controlled by
\begin{equation}
\mu_\chi^2=\sqrt{\frac{\alpha^2+\beta^2}{3}}.
\label{eq:phase-muchi}
\end{equation}
Therefore the diagonal channel $\chi$ exists whenever $(\alpha,\beta)\neq(0,0)$, since the $(3,8)$ sector always reorganizes into a conjugate pair of $i$-particle modes. By contrast, the off-diagonal replica-like channel survives only when
\begin{equation}
\mu_+^2=\mu_-^2,
\end{equation}
which is equivalent to
\begin{equation}
\beta=0.
\label{eq:phase-beta0}
\end{equation}

This leads to three qualitatively distinct loci.

\subsection*{1. The generic Cartan region: only the diagonal physical channel survives}

For generic points in the Cartan plane,
\begin{equation}
\beta\neq0,
\qquad
(\alpha,\beta)\neq(0,0),
\label{eq:generic-region}
\end{equation}
the off-diagonal subsectors $(4,5)$ and $(6,7)$ are no longer conjugate to each other, since in general $\mu_+^2\neq\mu_-^2$. As shown in Sect.~\ref{sec:operators}, the would-be off-diagonal neutral bilinear then develops a complex threshold,
\begin{equation}
\tau_{\mathrm{off}}
=
2\sqrt{\mu_+^2\mu_-^2}
+
i(\mu_+^2-\mu_-^2),
\label{eq:phase-tauoff}
\end{equation}
and the standard K\"all\'en--Lehmann interpretation of the replica-like channel is lost. In this region, the only surviving physical channel of the $i$-particle type is the diagonal one,
\begin{equation}
\chi
=
(\partial_\mu U_\nu-\partial_\nu U_\mu)
(\partial_\mu V_\nu-\partial_\nu V_\mu),
\end{equation}
with threshold
\begin{equation}
\tau_\chi=2\mu_\chi^2=2\sqrt{\frac{\alpha^2+\beta^2}{3}}.
\label{eq:phase-tauchi}
\end{equation}

Thus the generic Cartan deformation preserves a physical channel, but not the replica-like coexistence found in the original model.

\subsection*{2. The distinguished line $\beta=0$: coexistence of the two physical channels}

On the special line
\begin{equation}
\beta=0,
\qquad
\alpha>0,
\label{eq:replica-line}
\end{equation}
one has
\begin{equation}
\mu_+^2=\mu_-^2=\frac{\alpha}{2},
\qquad
\mu_\chi^2=\frac{|\alpha|}{\sqrt{3}}.
\label{eq:replica-line-masses}
\end{equation}
The off-diagonal sectors then form a genuine conjugate pair, and the neutral bilinear
\begin{equation}
\phi=\Phi_1=X^+Y^+ + X^-Y^-,
\end{equation}
regains the standard physical threshold
\begin{equation}
\tau_\phi=\alpha,
\label{eq:tauphi-replica-line}
\end{equation}
while the diagonal channel remains present with
\begin{equation}
\tau_\chi=\frac{2|\alpha|}{\sqrt{3}}.
\label{eq:tauchi-replica-line}
\end{equation}
Therefore the line $\beta=0$ is singled out as the locus where the model displays the simultaneous existence of both physical channels: the diagonal one and the off-diagonal replica-like one. This is precisely the situation realized by the original color-3 source.

In this sense, the color-3 model is not just a convenient example inside a larger family. It is the unique one-parameter subfamily in the Cartan plane for which the replica interpretation survives in the composite sector.

\subsection*{3. Degenerate loci and the undeformed point}

There are also special loci where one of the off-diagonal mass parameters vanishes,
\begin{equation}
\mu_+^2=0
\qquad\Longleftrightarrow\qquad
\alpha=\frac{\beta}{\sqrt{3}},
\label{eq:mupluszero}
\end{equation}
and
\begin{equation}
\mu_-^2=0
\qquad\Longleftrightarrow\qquad
\alpha=-\frac{\beta}{\sqrt{3}}.
\label{eq:muminuszero}
\end{equation}
Along these lines, one off-diagonal subsector becomes massless while the other remains nontrivial. The standard off-diagonal $i$-particle construction degenerates there, and the replica-like interpretation is again absent.
This should not be interpreted as a restoration of the replica-like structure or of a full embedded $SU(2)$ symmetry; rather, it reflects the fact that the Cartan source becomes orthogonal to one of the off-diagonal root directions, so that the corresponding complex shift vanishes at quadratic order.

Finally, at the origin
\begin{equation}
\alpha=\beta=0,
\label{eq:origin}
\end{equation}
the soft breaking disappears, the quadratic form reduces to the ordinary Landau-gauge Yang--Mills one, and the $i$-particle splitting is absent. In this limit neither of the channels discussed above should be regarded as isolated by the present mechanism.

\subsection*{4. Geometric interpretation}

The phase structure may therefore be summarized geometrically as follows:

\begin{itemize}
\item for generic $(\alpha,\beta)$ away from the origin, the model possesses only the diagonal physical channel $\chi$;
\item on the distinguished line $\beta=0$, the off-diagonal replica-like channel $\phi$ reappears and coexists with $\chi$;
\item on the lines $\alpha=\pm\beta/\sqrt{3}$, one of the off-diagonal sectors becomes degenerate;
\item at the origin, the soft-breaking mechanism is switched off.
\end{itemize}

The main conclusion is that the space of admissible physical composite operators is not constant over the Cartan plane. The orientation of the source does not merely deform pole positions: it determines which neutral channels retain a physical spectral interpretation. This is the sense in which the Cartan plane acquires a genuine phase structure already at the level of quadratic propagation and one-loop composite operators.

\section{Conclusions}
\label{sec:conclusions}

In this work we revisited the explicitly broken $SU(3)$ Yang--Mills model introduced in \cite{Amaral:2013ReplicaSU3} and extended it from the special color-3 source to a generic Cartan-valued soft source. Our aim was not merely to deform the old construction algebraically, but to understand which of its physically relevant features persist once the source is allowed to explore the full Cartan plane.

The first result is that the quadratic theory admits a natural and transparent description for a generic source $(\alpha,\beta)$. The off-diagonal sectors $(4,5)$ and $(6,7)$ acquire distinct complex mass parameters controlled by different linear combinations of $\alpha$ and $\beta$, while the $(3,8)$ block continues to diagonalize into a conjugate pair of $i$-particle modes. In this sense, the generic Cartan deformation preserves a confining-type complex-pole structure, but does so in a way that is no longer symmetric between the two off-diagonal subsectors.

The second and more important result concerns the composite sector. We showed that the survival of neutral composite operators with a standard K\"all\'en--Lehmann interpretation imposes a much stronger constraint than simple algebraic neutrality or Hermiticity. The diagonal channel built from the normal modes of the $(3,8)$ block survives everywhere away from the undeformed origin. By contrast, the off-diagonal replica-like channel survives only on the special locus $\beta=0$, where the $(4,5)$ and $(6,7)$ sectors remain a genuine conjugate pair. Therefore, the original color-3 model is singled out dynamically inside the Cartan plane: it is precisely the locus for which the two physical channels coexist.

This gives a new interpretation to the original construction. The color-3 source is not merely one convenient choice among many possible Cartan directions. Rather, it is the unique one-parameter direction for which the replica-like physical channel is compatible with the standard $i$-particle mechanism at one loop. Put differently, once one asks for the survival of the physical composite sector, the Cartan plane acquires a nontrivial phase structure.

The picture that emerges is thus the following. For generic Cartan deformations, the model retains a diagonal physical channel but loses the off-diagonal replica-like one. On the distinguished line $\beta=0$, both channels are present. This provides a simple physical selection principle for the orientation of the soft source and suggests that the admissible composite sector carries information that is not visible from the pole structure alone.

This unique physical phase naturally raises the question of its dynamical origin. In Appendix \ref{app:ssb}, we address this by presenting a BRST-invariant spontaneous symmetry breaking mechanism that generates the Cartan soft source. This formulation not only provides a natural dynamical realization of the deformation studied in the main text but also proves the strict algebraic renormalizability and Ward-identity structure of the theory.

While the quadratic theory and its one-loop composite sector are now algebraically and dynamically well-defined, some questions remain open. In particular, it would be interesting to investigate whether the phase structure identified here persists beyond quadratic order and how it interfaces with more realistic glueball-like operators and with the broader problem of analyticity in confining gauge theories.

Our results indicate that the study of explicit soft breakings should not be organized only around which propagators exhibit complex poles, but also around which neutral composite channels retain a physical spectral meaning. From this perspective, the Cartan plane provides a natural laboratory in which confinement-inspired propagator structures, residual symmetry, and operator analyticity can be studied in a unified way.

It would be interesting to investigate whether traces of this Cartan phase structure can be identified in nonperturbative studies of neutral composite correlators, for instance in lattice-inspired analyses. From that perspective, the present results suggest that the emergence of an admissible glueball-like sector may depend not only on the positivity properties of the elementary propagators, but also on the effective orientation of the soft breaking in Cartan space.

\appendix

\section{A spontaneous realization of the Cartan soft source and Algebraic Renormalization}
\label{app:ssb}

In the main text, the Cartan-valued source $\langle J^a\rangle=\alpha\,\delta^{a3}+\beta\,\delta^{a8}$ was treated as a physical soft insertion. It is natural to ask whether such a source can arise dynamically from a BRST-invariant sector without relying on an explicit breaking, and whether this construction is algebraically renormalizable. In this appendix, we present a spontaneous realization of this source and provide the complete functional setup for its ultraviolet stability.

\subsection{The Spontaneous Mechanism}

We introduce an adjoint BRST doublet $(\epsilon^a,\varphi^a)$ with transformations
\begin{equation}
s\,\epsilon^a=\varphi^a,
\qquad
s\,\varphi^a=0,
\label{eq:brst-doublet-app}
\end{equation}
where $\epsilon^a$ is Grassmann odd and $\varphi^a$ is Grassmann even. We utilize this doublet to introduce the dimension-two operator $d^{abc}A_\mu^bA_\mu^c$ not via an external source, but coupled to a field that can acquire a mass through spontaneous symmetry breaking. 

We define the BRST-exact auxiliary action
\begin{equation}
S_{\mathrm{aux}}
=
s\int d^4x
\left[
a_1 M\,\epsilon^a\,\partial_\mu A_\mu^a
+
\frac{a M}{2}\,\epsilon^a d^{abc}A_\mu^bA_\mu^c
\right],
\label{eq:Saux-app}
\end{equation}
where $a$ and $a_1$ are dimensionless parameters and $M$ is an auxiliary mass scale. The first term provides a necessary kinetic-like structure for the $\epsilon^a$ field. Applying the nilpotent BRST operator $s$, with $s A_\mu^a = -(D_\mu c)^a$, and following the graded Leibniz rule $s(\epsilon^a X) = \varphi^a X - \epsilon^a (sX)$, the expansion yields:
\begin{equation}
\begin{split}
S_{\mathrm{aux}}
=
\int d^4x
\Bigg[&
a_1 M\,\varphi^a\,\partial_\mu A_\mu^a
+
a_1 M\,\epsilon^a\,\partial_\mu(D_\mu c)^a
\\
&
+ \frac{a M}{2}\,\varphi^a d^{abc}A_\mu^bA_\mu^c
+ 
a M\,\epsilon^a d^{abc}A_\mu^b(D_\mu c)^c
\Bigg].
\end{split}
\label{eq:Saux-expanded-app}
\end{equation}

To allow for spontaneous symmetry breaking, we supplement the action with a renormalizable real scalar sector,
\begin{equation}
S_{\varphi}
=
\int d^4x
\left[
\frac12\,\partial_\mu\varphi^a\,\partial_\mu\varphi^a
-
\frac{M^2}{2}\,\varphi^a\varphi^a
+
\frac{\lambda}{4}\,(\varphi^a\varphi^a)^2
\right].
\label{eq:Sphi-app}
\end{equation}
The potential is minimized at
\begin{equation}
\langle\varphi^a\varphi^a\rangle=\nu^2,
\qquad
\nu^2=\frac{M^2}{\lambda}.
\label{eq:vev-cond-app}
\end{equation}
We can therefore shift the scalar field around a chosen vacuum direction $u^a$ (with $u^a u^a = 1$):
\begin{equation}
\varphi^a=\widetilde{\varphi}^a+\nu\,u^a.
\label{eq:shift-app}
\end{equation}
Substituting (\ref{eq:shift-app}) into the third term of (\ref{eq:Saux-expanded-app}), it splits into
\begin{equation}
\frac{a M}{2}\,\widetilde{\varphi}^a d^{abc}A_\mu^bA_\mu^c
+
\frac{a M \nu}{2}\,u^a d^{abc}A_\mu^bA_\mu^c.
\label{eq:generated-soft-app}
\end{equation}
The vacuum dynamically generates an effective source $\langle J^a \rangle = a M \nu u^a$. By orienting $u^a$ in the Cartan plane, we recover the precise two-parameter deformation $(\alpha,\beta)$ studied in the main text without any explicit symmetry breaking.

\subsection{Symmetries Compatible with the Quantum Action Principle}

To establish the stability of the action in the ultraviolet, we must analyze the symmetries that can be written in functional form respecting the Quantum Action Principle (QAP). We introduce the external source sector to properly quantize the non-Abelian symmetries:
\begin{equation}
S_{\mathrm{ext}}=\int d^{4}x \left\{ -\Omega^{a}_{\mu}(D_{\mu}c)^{a} + \frac{g}{2}f^{abc}L^a c^{b}c^{c}\right\}.
\end{equation}
The complete action is defined as $\Sigma = S_{\mathrm{YM}} + S_{\mathrm{gf}} + S_{\mathrm{aux}} + S_{\varphi} + S_{\mathrm{ext}}$. 

This action satisfies the Landau gauge fixing and anti-ghost equations:
\begin{eqnarray}
\frac{\delta \Sigma}{\delta b^{a}}&=&i\partial_{\mu}A_{\mu}^{a}\nonumber \\
\bar{G}^a(\Sigma) &=& \partial_{\mu}\frac{\delta \Sigma}{\delta \Omega_{\mu}^{a}} + \frac{\delta \Sigma}{\delta \bar{c}^{a}}=0.
\end{eqnarray}
It obeys the Slavnov-Taylor identity:
\begin{eqnarray}
\mathcal{S}(\Sigma) &=& \int d^{4}x  \left\{ \frac{\delta \Sigma}{\delta A_{\mu}^{a}}\frac{\delta \Sigma}{\delta \Omega_{\mu}^{a}} +
\frac{\delta \Sigma}{\delta c^{a}}\frac{\delta \Sigma}{\delta L^{a}} +\varphi^{a}\frac{\delta \Sigma}{\delta\epsilon^{a}} + ib^{a}\frac{\delta \Sigma}{\delta\bar{c}^{a}}\right\} = 0,
\end{eqnarray}
and the Ghost equation (which receives a linear breaking typical of the Landau gauge):
\begin{eqnarray}
G^{a}(\Sigma)&=& \int d^{4}x \left\{ \frac{\delta \Sigma}{\delta c^{a} } -igf^{abc}( \bar{c}^{b}+a_{1}M\epsilon^{b})\frac{\delta \Sigma}{\delta b^{c} } \right\} \nonumber \\
&=& \int d^{4}x\,  gf^{abc} \left\{ \Omega_{\mu}^{b}A_{\mu}^{c}- L^{b}c^{c} \right\}.
\end{eqnarray}

The presence of the $d^{abc}$ operator requires specific functional identities. We identify the following constraints:
\begin{eqnarray}
E^{a}(\Sigma) &=& \int d^{4}x \left\{ \frac{\delta \Sigma}{\delta \epsilon^{a}} - 2iaM d^{abc} c^{b} \frac{\delta\Sigma}{\delta b^{c}}\right\} = 0, \\
W^{a}(\Sigma) &=& \int d^{4}x \left\{gf^{abc}\theta^{b}\frac{\delta \Sigma}{\delta \theta^{c}}-i g a_{1}M f^{abc} \varphi^{b}\frac{\delta \Sigma}{\delta b^{c}}\right\} = 0,
\end{eqnarray}
where $\theta^a$ denotes all fields and sources except the introduced doublet. Finally, we have two further relations:
\begin{eqnarray}
R(\Sigma)&=&  \int d^{4}x \left\{ c^{a}\frac{\delta\Sigma}{\delta \bar{c}^{a}} - i \frac{\delta\Sigma}{\delta L^{a}}\frac{\delta\Sigma}{\delta b^{a}}\right\} =0, \\
R^{a}(\Sigma)&=& \int d^{4}x\, d^{abc}\left\{ c^{b}\frac{\delta\Sigma}{\delta \bar{c}^{c}} - i \frac{\delta\Sigma}{\delta L^{b}}\frac{\delta\Sigma}{\delta b^{c}}\right\} =0.
\end{eqnarray}
While the first equation is a standard rigid identity related to the $SL(2,\mathbb{R})$ symmetry of the ghost sector in the Landau gauge, the second is a novel, model-specific constraint. It arises strictly due to the completely symmetric $d^{abc}$ coupling of the spontaneous breaking mechanism and plays a crucial role in restricting the admissible counterterms to the exact Cartan orientation.

\subsection{Renormalization and Necessary Relations}

To construct the counterterm action, we introduce the nilpotent linearized Slavnov-Taylor operator $\mathcal{B}_\Sigma$ (where $\mathcal{B}_\Sigma^2 = 0$), which extends the BRST symmetry to the sources:
\begin{equation}
\begin{split}
\mathcal{B}_\Sigma = \int d^{4}x \Bigg\{ & \frac{\delta \Sigma}{\delta A_{\mu}^{a}}\frac{\delta }{\delta \Omega_{\mu}^{a}} + \frac{\delta \Sigma}{\delta \Omega_{\mu}^{a}}\frac{\delta }{\delta A_{\mu}^{a}} + 
\frac{\delta \Sigma}{\delta c^{a}}\frac{\delta }{\delta L^{a}} \\
& + \frac{\delta \Sigma}{\delta L^{a}}\frac{\delta }{\delta c^{a}} + \varphi^{a}\frac{\delta }{\delta\epsilon^{a}} + ib^{a}\frac{\delta }{\delta\bar{c}^{a}} \Bigg\}.
\end{split}
\end{equation}
The commutation relations with the functional operators define the consistency of the algebraic setup:
\begin{eqnarray}
 \left[\frac{\delta }{\delta b^{a}},\mathcal{B}_\Sigma \right] &=& i\bar{G}^{a}, \qquad \{ G^{a},\mathcal{B}_\Sigma \} = W^{a}, \qquad \{E^{a},\mathcal{B}_\Sigma\} = 2aM \mathcal{R}^{a},
\end{eqnarray}
where $G^a, W^a, E^a$, and $\mathcal{R}^a$ denote the linear functional operators corresponding to the identities of the previous section.

Given a multiplicative renormalization, the quantum action $\Gamma = \Sigma + \hbar \Gamma^{c}$ must satisfy $\mathcal{B}_\Sigma\Gamma^c = 0$. The counterterm action can be decomposed as $\Gamma^{c} = \Gamma^{0} + \mathcal{B}_\Sigma \Gamma^{-1}$, where $\Gamma^{0}$ is the cohomological sector and $\Gamma^{-1}$ is a functional of dimension 4 and ghost number $-1$. Applying all symmetries yields the most general functional:
\begin{eqnarray}
\Gamma^{-1}&=& \int d^{4}x \Big\{  y_{1} (\partial_{\mu}\bar{c}^{a} + \Omega_{\mu}^{a}) A_{\mu}^{a} + y_{1} \frac{a M}{2} d^{abc}\epsilon^a A_{\mu}^{b}A_{\mu}^{c}+ y_{2}\partial_{\mu}\epsilon^{a}\partial_{\mu}\varphi^{a} \nonumber \\
&& \qquad \quad + y_{3}M^{2}\epsilon^{a}\varphi^{a} + y_{4}\lambda \epsilon^{a}\varphi^a (\varphi^{b} \varphi^{b})  + y_{5}\epsilon^{a}\partial_{\mu}A_{\mu}^{a} \Big\}.
\end{eqnarray}
Applying the $\mathcal{B}_\Sigma$ operator, we obtain the full counterterm action:
\begin{eqnarray}
\Gamma^{c} &=& \int d^{4}x \Big\{ \frac{y_{0}}{4} F_{\mu\nu}^a F_{\mu\nu}^a  -y_{1}A_{\mu}^{a}(D_{\nu}F_{\nu\mu})^{a} + y_1 (\partial_{\mu}\bar{c}^{a} + \Omega_{\mu}^{a})\partial_{\mu}c^{a} \nonumber \\
&& \qquad \quad +y_{1}aM d^{abc}(\varphi^{a}A_{\mu}^{b}A_{\mu}^{c}+ \epsilon^{a}A_{\mu}^{b}\partial_{\mu}c^{c})+y_{2}\partial_{\mu}\varphi^{a}\partial_{\mu}\varphi^{a} \nonumber \\
&& \qquad \quad + y_{3}M^{2}\varphi^{a}\varphi^{a} + y_{4}\lambda (\varphi^{a}\varphi^{a} )^{2} \Big\}.
\end{eqnarray}

We define the multiplicative renormalization for the fields and parameters as $\Phi' \rightarrow (1 + \hbar z_\Phi) \Phi$. Matching $\Sigma(\Phi') = \Sigma(\Phi) + \hbar \Gamma^c(\Phi)$ completely determines the renormalization constants in terms of the independent coefficients $y_i$:
\begin{eqnarray}
z_{A}&=&y_{0}+ 2y_{1},\qquad z_{g}=-\frac{y_{0}}{2},\qquad z_{\varphi}=\frac{y_{2}}{2},\qquad z_{\epsilon}=y_{1}+\frac{y_{2}}{2}\nonumber \\
z_{m}&=& y_{3}-\frac{y_{2}}{2},\qquad z_{\lambda}=y_{4}-2y_{2},\qquad z_{a}=-\left(\frac{y_{0}}{2}+y_{3}\right),\qquad z_{a_{1}}=-y_{3}\nonumber \\
z_{L}&=& -2z_{c},\qquad z_{\Omega}=z_{\bar{c}}=0.
\end{eqnarray}
The complete closure of the $Z$-factors demonstrates that the spontaneous symmetry breaking mechanism generating the Cartan source is fully consistent and algebraically renormalizable.

\end{document}